\documentclass[sn-mathphys,Numbered]{sn-jnl}

\usepackage{graphicx}%
\usepackage{multirow}%
\usepackage{amsmath,amssymb,amsfonts}%
\usepackage{amsthm}%
\usepackage{mathrsfs}%
\usepackage[title]{appendix}%
\usepackage{xcolor}%
\usepackage{textcomp}%
\usepackage{manyfoot}%
\usepackage{booktabs}%
\usepackage{algorithm}%
\usepackage{algorithmicx}%
\usepackage{algpseudocode}%
\usepackage{listings}%

\theoremstyle{thmstyleone}%
\theoremstyle{thmstyletwo}%

\theoremstyle{thmstylethree}%

\begin{document}

\title[Article Title]{Neutron star calculations with the phenomenological three-nucleon force}

\author*{\fnm{H.} \sur{Moeini}}\email{h.moeini@shirazu.ac.ir}

\author*{\fnm{G.H.} \sur{Bordbar}}\email{ghbordbar@shirazu.ac.ir}

\affil{\orgdiv{Department of Physics}, \orgname{School of Science,
Shiraz University}, \orgaddress{\city{Shiraz}, \postcode{71454}, \country{Iran}}}

\abstract{In this work, we have studied the effect of three-nucleon interaction on the neutron stars structure.
In our calculations, we have considered the neutron star matter as a beta-stable nuclear matter.
We have put the results concerning the TBF effect in perspective against two-body results and other calculations of three-nucleon interactions, 
using the Urbana $\it{v_{14}}$ potential and the parabolic approximation of the nuclear-matter energy for approximating the problem of asymmetric nuclear matter. 
As such, solving the Tolman-Oppenheimer-Volkoff equation, we have estimated bulk properties of neutron stars 
and investigated how the present calculations would agree with the expected dynamical-stability condition.}



\maketitle

\section{Introduction}
\label{sec:intro}
The notion of introducing three-nucleon forces (TBF) has manifested itself to be indispensable in deriving bulk properties of symmetric nuclear matter such as the saturation density, energy, symmetry energy, and incompressibility -- the interest to the latter of which concerns the physics of neutron stars and evolution of supernovae. The TBF effect in the equation of state (EOS) of high-density nuclear matter is envisaged to be substantial and, as such, vital in addressing high-energy heavy-ion collisions and properties of dense objects such as neutron stars~\cite{Kievsky2010,Lovato2012,WiringaFiks1988}. As the maximum mass of such objects is known to depend sensitively on EOS~\cite{BordbarHayati2006}, their bulk properties such as radius at maximum mass can thus be influenced by TBF. {This is especially the case at high nuclear-matter densities where there are also a lot of interest in, for instance, modified gravities to study the astrophysical dynamics, matter instability and singularities appearing in collapse processes of compact objects~\cite{Bamba2011,Yousaf2016,Bhatti2023}.} Thus, neutron stars can be viewd as astrophysical laboratories to test nuclear matter EOS at high densities, since recent discoveries of about 1.97~\cite{Demorest2010}, 2.01~\cite{Antoniadis2013}, 2.10~\cite{Cromartie2020}, and 2.3~$M_{\odot}$~\cite{Linares2018} neutron stars -- which are heavier than most of the observed ones in binary systems of $1.2-1.6~M_{\odot}$ ~\cite{Ozel2012,Foucart2020} -- have challenged many of the EOS models.

{By predicting a greater burst of compact objects -- which have profound significance for experimental astrophysics -- modified gravity theories stand out in favoring the existence of super-massive structures of smaller radii than foreseen by general relativity. These theories provide a framework for describing also the distribution of compact objects, employing an equation of state within their own context~\cite{Bhatti2024,Bhatti2022}. Hence, it is imperative that gravity theories with suitable frameworks could address the effects of mass, EOS parameters, and electric charge -- within the largest ranges of possible values -- that could fulfill the stability requirements~\cite{Yousaf2023}. It is important to have suitable frameworks that would allow for searching models which could present a smooth matching between two different space-times at a separation hypersurface of compact objects, such as isotropic perfect fluid stars, supported by thin shells in modified gravity~\cite{Bhatti2023_}. As such, one could derive among others surface energy densities as well as various ingredients of surface pressures at separation hypersurface~\cite{Bhatti2024}.}

The lowest order constrained variational method (LOCV) was established for $v_8$~\cite{LagarisPandharipande1980}, $v_{12}$ and Urbana $v_{14}$ (U$v_{14}$)~\cite{LagarisPandharipande1981}, Argonne $v_{14}$ (A$v_{14}$)~\cite{WiringaSmith1984}, and A$v_{18}$~\cite{WiringaStokes1995} potentials and has delivered comparable results to variational methods that incorporate many-body contributions~\cite{BordbarModarres1998}.
Using LOCV, we have studied bulk properties of symmetric nuclear and pure neutron matter \cite{BordbarModarres1997,Bordbar2003,BigdeliBordbarRezaei2009,RezaeiBordbar2017,BordbarRezaei2013,BordbarBigdeli2007,BordbarBigdeli2008,BordbarRezaeiMontakhab2011,RezaeiBigdeliBordbar2015} as well as asymmetric nuclear matter~\cite{BordbarModarres1998,ModarresBordbar1998,BordbarBigdeli2008_2nd,BordbarBigdeli2007_2nd,BigdeliBordbarPoostforush2010}, especially in connection with neutron star properties~\cite{BordbarHendi2016,RezaeiBordbar2016,HendiBordbar2017,EslamPanahBordbar2017,EslamPanahYazdizadeh2019,BordbarKarami2022}. It should be stated that, in what follows, what we refer to as the TBF effect is specifically assumed to be the combined effects of a two-pion-exchange potential and a phenomenological repulsive term~\cite{Carlson1983,Pudliner1995}.

Similar to other potentials, since the fitted U$v_{14}$ or A$v_{14}$ (hereafter, referred to as UA$\it{v_{14}}$) to two-nucleon data underestimates binding energies of light nuclei (like $^3$H and $^4$He) and at the same time overbinds nuclear matter, a three-body term is introduced to take into account the required binding adjustments and also the theoretical anticipation of the existence of non-nucleonic resonances like $\Delta$ states, which are overlooked in building up two-nucleon potentials~\cite{WiringaFiks1988}. Previously, we have reported on the symmetric nuclear matter calculations within the LOCV framework employing UA$\it{v_{14}}$ potentials and accounting for the phenomenological TBF effect based upon the UVII three-nucleon potential~\cite{MoeiniBordbar2022}. Our U$\it{v_{14}}$ calculations resulted in closer values of saturation energy, incompressibility, and symmetry energy to the empirical values than the A$\it{v_{14}}$ results did. As such, we have presented here our results using U$\it{v_{14}}$ and investigated the TBF effect on the pure neutron and beta-stable matter and, hence, on neutron stars purely made out of nucleons. In this regard, a parabolic approximation of the energy of asymmetric matter was employed to derive EOS.

In what follows, we first present a short review of the zero-temperature two- and three-nucleon interactions and energy contributions in the UA models, using the correlation functions derived within the LOCV formalism. Next, we provide an overview of the beta-stability condition and how the bulk properties of a beta-stable neutron star were derived under the assumption of hydrostatic equilibrium formulated within general relativity by the TOV equation \cite{Tolman1934,Tolman1939,OppenheimerVolkoff1939}. Finally, we present the results and conclusions.

%%%%%%%%%%%%%%%%%%%%%%%%%%%%%%%%%%%%%%%%%%%%%%%%%%%%%%%%%%%%%%%%%%%%%%%%%%%%%

\section{Two- and three-nucleon interactions}
\label{sec:InterNucleonInteractions}
Below pion-production energies, the low-energy Hamiltonian can be approximated by taking into account only two- and three-body terms as \cite{NATOASIseries1997}:
\begin{equation}
H=-\sum_{{i\leq A}}\frac{\hbar^2}{2m}\nabla^2_i+\sum_{{i< j\leq A}}V_{ij}+\sum_{{i< j< k\leq A}}V_{ijk}.
\label{Eq:Hamiltonian}
\end{equation}
where $V_{ij}$ and $V_{ijk}$ stand for two-body and three-body potentials, respectively. The two-body potential, constrained by $NN$ scattering data, is constructed in the UA$\it{v_{14}}$ models on the basis of fourteen operators ($O_{12}$) and takes the following form~\cite{LagarisPandharipande1981}:
\begin{eqnarray}
V(12)=\sum_{p=1}^{14}v^{(p)}(r_{12})O^{(p)}_{12}
\label{v12}
\end{eqnarray}

The three-body potential is assumed to be comprised of a phenomenological medium-range repulsive term $V_{ijk}^R$ and a long-range attractive term corresponding to two-pion exchange $V_{ijk}^{2\pi}$ as follows \cite{FujitaMiyazawa1957,GibsonMcKellar1988,WiringaFiks1988,LagarisPandharipande1981_2}:
\begin{eqnarray}
V_{ijk}=V_{ijk}^R+V_{ijk}^{2\pi}=U\sum_{cyc}T^2_{\pi}(r_{ij})T^2_{\pi}(r_{ik})+\nonumber\\
A_{2\pi}\sum_{cyc}\Big(\{X_{ij}^{\pi},X_{ik}^{\pi}\}\{{\boldsymbol\tau}_i\cdot{\boldsymbol\tau}_j,
{\boldsymbol\tau}_i\cdot{\boldsymbol\tau}_k\}+\nonumber\\
\frac{1}{4}[X_{ij}^{\pi},X_{ik}^{\pi}][{\boldsymbol\tau}_i\cdot{\boldsymbol\tau}_j,{\boldsymbol\tau}_i\cdot{\boldsymbol\tau}_k]\Big)
\label{Eq:A2pi}
\end{eqnarray}
where
\begin{eqnarray}
X_{ij}^{\pi}&=&Y_{\pi}(r_{ij}){\boldsymbol\sigma}_i\cdot{\boldsymbol\sigma}_j+T_{\pi}(r_{ij})\textbf{S}_{ij}, \nonumber\\
Y_{\pi}(r)&=&\frac{e^{-m_{\pi}r}}{m_{\pi}r}\big(1-e^{-cr^2}\big),\nonumber\\
T_{\pi}(r)&=&\Big(1+\frac{3}{m_{\pi}r}+\frac{3}{m_{\pi}^2r^2}\Big)Y_{\pi}(r)\big(1-e^{-cr^2}\big),\nonumber\\
\textbf{S}_{ij}&=&3({\boldsymbol\sigma}_i\cdot\hat{\textbf{r}}_{ij})({\boldsymbol\sigma}_j\cdot\hat{\textbf{r}}_{ij})-{\boldsymbol\sigma}_i\cdot{\boldsymbol\sigma}_j,
\label{Eq:Xij}
\end{eqnarray}
the details of which, including calculation of the constants $A_{2\pi}=-0.0331$~MeV and $U=0.0045$~MeV for the U$v_{14}$ potential as well as the two- and three-body nucleon-nucleon energy contributions, were presented in~\cite{MoeiniBordbar2022}. As such, inter-particle interactions were accounted for by employing inter-nucleon correlation functions $f(ij)$ calculated within the LOCV formalism~\cite{BordbarModarres1998}. Hence, the expectation value of the three-nucleon interaction was shown to relate to the three-body radial distribution function defined as~\cite{Clark1979}:
\begin{eqnarray}
g(\textbf{r}_1,\textbf{r}_2,\textbf{r}_3)=f^2(r_{12})f^2(r_{23})f^2(r_{13})g_{F}(\textbf{r}_1,\textbf{r}_2,\textbf{r}_3)
\label{gFunction}
\end{eqnarray}
in which $g_{F}(\textbf{r}_1,\textbf{r}_2,\textbf{r}_3)$ is the so-called three-body radial distribution function for the ground state of the interaction-free Fermi-gas.

It should be noted that in our previous work, using LOCV in conjunction with different two-body potentials, we had investigated the EOS of nuclear matter in presence of the three-nucleon interaction (TNI)~\cite{BordbarModarres1997}. Here, the effect of TNI plus U$v_{14}$ is but an approximation of the effect of $V_{ijk}$ in which TNI is assumed to be composed of repulsive (TNR) and attractive (TNA) terms -- accounting for the effects of $l=0$ and $l\neq0$, respectively. The three-nucleon repulsion term is assumed as an exponential term $e^{-\gamma\rho}$ multiplied by the intermediate-range part of $V(12)$, namely $v_I^{(p)}(r_{12})$. The exponential term is introduced to also approximate higher-than-third order interactions, where the third-order interactions correspond to $-\gamma\rho v_I^{(p)}(r_{12})$ terms with more complicated spin-isospin dependence than $V_{ijk}^R$~\cite{LagarisPandharipande1981,WiringaFiks1988}.
%%%%%%%%%%%%%%%%%%%%%%%%%%%%%%%%%%%%%%%%%%%%%%%%%%%%%%%%%%%%%%%%%%%%%%%%%%%%%%%%%%%%%

\section{Beta-stable matter and the neutron star calculations}
\label{sec:betaStableEqStateNeutStar}
As the EOS of nucleonic matter is expected to either govern or have direct influence in bulk properties of the neutron star, we briefly lay out the framework for such envisaged connection between microscopic EOS and neutron star's bulk properties like its maximum mass. We shall employ the U$\it{v_{14}}$ potential in conjunction with a three-body contribution, calculated based upon the phenomenological UVII model addressed in Sec.~\ref{sec:InterNucleonInteractions}.
The beta-stability condition requires the inclusion of leptonic relativistic contributions to the energy content of the neutron star: 
\begin{equation}
E_{lep}= \sum_i \frac{{m_i}^4c^5}{8{\pi}^2{\hbar}^3\rho}\Bigg(\frac{\hbar k_i}{m_i c}\Big[1+\Big(\frac{\hbar k_i}{m_i c}\Big)^2\Big]^{1/2}\Big[1+2\Big(\frac{\hbar k_i}{m_i c}\Big)^2\Big]-\sinh^{-1}\Big(\frac{\hbar k_i}{m_i c}\Big)\Bigg)
\label{rho3g}
\end{equation}
in which $i$ runs over electrons and muons, and $k_i$ represents their respective Fermi momenta, which are related as dictated by the following beta-stability condition:
\begin{equation}
\mu_n=\mu_p+\mu_e=\mu_p+\mu_{\mu}
\end{equation}
in which $\mu_j$ {(in MeV)} stands for the chemical potential of neutrons, protons, electrons, or muons. Hence, knowing that $\rho=\rho_p+\rho_n$ {(in fm$^{-3}$)} and assuming the charge neutrality condition $\rho_p=\rho_e+\rho_{\mu}$ for relativistic electrons and muons with chemical potentials of approximately $\hbar c\big(3\pi^2\rho_{e,\mu}\big)^{1/3}$, we used the parabolic approximation for the energy of asymmetric matter \cite{LagarisPandharipande1981_3}:
\begin{eqnarray}
E(\rho,\rho_p)=\frac{3}{5}\frac{\hbar^2}{2m_N}\big(3\pi^2\rho\big)^{2/3}\Big[(\rho_p/\rho)^{5/3}+\nonumber\\
(1-\rho_p/\rho)^{5/3}\Big]+V_0(\rho)+(1-2\rho_p/\rho)^2 E_{symm}(\rho)
\label{asymMatEnergy}
\end{eqnarray}
in which the first term is the Fermi-gas kinetic energy $T_F(\rho,\rho_p/\rho)$ -- with  $T_F(\rho,\rho_p/\rho)+V_0(\rho)$ representing the symmetric nuclear-matter energy -- resulting in the following relation to be used in conjunction with the above relations for extracting the nucleonic and leptonic densities of beta-stable matter:
\begin{eqnarray}
\mu_n-\mu_p=\frac{\hbar^2}{2m_N}\big(3\pi^2\rho\big)^{2/3}\Big[\big(1-\rho_p/\rho\big)^{2/3}-\nonumber\\\big(\rho_p/\rho\big)^{2/3}\Big]+4\big(1-2\rho_p/\rho\big)E_{symm}(\rho)
\end{eqnarray}
$V_0(\rho)$ and the symmetry energy $E_{symm}(\rho)$ were obtained from the symmetric nuclear matter and pure neutron matter calculations, assuming the parabolic approximation.

Astrophysically, a star's equilibrium is reached owing to the balance between internal pressure and gravitational force. Such balance is expressed by an underlying hydrostatic equilibrium equation (HEE) established by Tolman, Oppenheimer, and Volkoff (TOV) within the framework of Einstein gravity. Using the TOV equation which holds for the general-relativistic hydrostatic equilibrium:
\begin{equation}
\frac{dP}{dr}=-\frac{G}{r^2}\Big[\epsilon(r)+P(r)/c^2\Big]\frac{m(r)+4\pi r^3 P(r)/c^2}{1-2G m(r)/rc^2},
\end{equation}
the bulk properties (mass and radius) of the beta-stable neutron star were thus calculated as a function of the central pressure $P_c$ {(in MeV/fm$^{3}$)} and mass density $\epsilon_c$ {(in gr/cm$^{3}$)}. Here, $G$, $\epsilon(r)=\rho\big[E/N(\rho)+m_N c^2\big]$, and $m(r)$ are, respectively, the gravitational constant, the mass density at distance $r$ from the center of the assumed spherical neutron star of radius $R$, and the total mass enclosed within a sphere of radius $r<R$. The neutron-star mass is thus $m(R)$ and $R$ is obtained by integrating the TOV equation from $r=0$ to $r=R$, at which point the pressure is assumed to vanish effectively (see \cite{Shapiro1983} for details).

%%%%%%%%%%%%%%%%%%%%%%%%%%%%%%%%%%%%%%%%%%%%%%%%%%%%%%%%%%%%%%%%%%%%%%%%%%%%%%%%%%%%%
\begin{figure*}[h]
\includegraphics[width=0.95\textwidth]{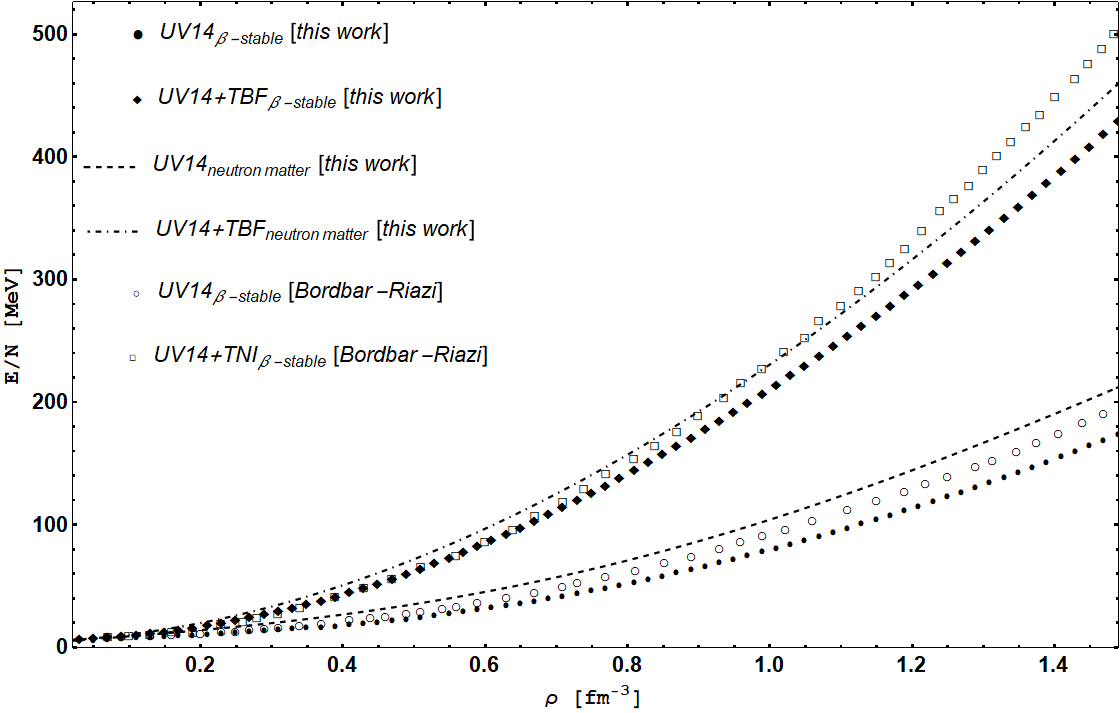}
\caption{\label{fig:eqStateBetaStable} Various U$\it{v_{14}}$ results for the binding energy per nucleon, as a function of nucleon density, of beta-stable as well as neutron matter in presence/absence of the TBF contribution. The data labeled as Bordbar-Riazi were extracted from ~\cite{BordbarRiazi2002}.}
\end{figure*}

\section{Results}
\label{sec:results}
\begin{figure*}[h]
\includegraphics[width=0.95\textwidth]{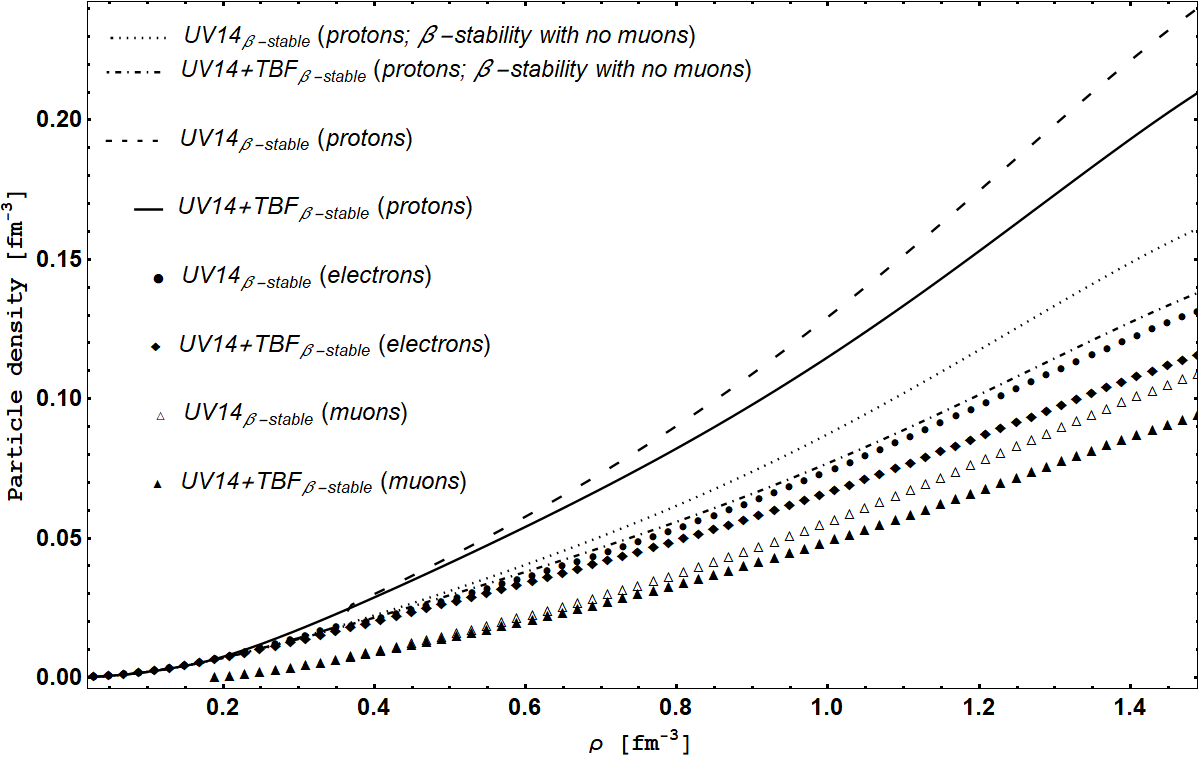}
\caption{\label{fig:particleDensities} Various particle densities, as a function of nucleon density, of beta-stable matter with and without the TBF contribution. The dotted and dash-dotted data represent the case in which the beta-stability equilibrium is only governed through $n\leftrightarrow p+e^-$.}
\end{figure*}

Fig.~\ref{fig:eqStateBetaStable} compares various U$\it{v_{14}}$ results for the mean binding energy of beta-stable as well as neutron matter in presence and absence of three-body contribution estimated as TBF or TNI. As such, our results for various particle densities of the beta-stable matter are shown in Fig.~\ref{fig:particleDensities} and the pressure, sound-velocity, and dynamical stability results are presented in Figs.~\ref{fig:pressure}, \ref{fig:soundVelocity}, and \ref{fig:AdiabaticIndex}. The results in Figs.~\ref{fig:Mass_Density} and \ref{fig:Mass_Radius} are derived based on the solutions of the TOV equation.

%%%%%%%%%%%%%%%%%%%%%%%%%%%%%%%%%%%%%%%%%%%%%%%%%%%%%%%%%%%%%%%%%%%%%%%%%%%%%%%%%%%%%

\subsection{Binding energy}
\label{results:1}
Our calculations, using U$\it{v_{14}}$ potential and introducing a TBF effect based on the phenomenological UVII model~\cite{MoeiniBordbar2022}, resulted in saturation density, incompressibility, and symmetry energy values of, respectively, about 0.364 (0.178)~fm$^{-3}$, 302 (193)~MeV, and 44.8 (29.2)~MeV, for U$\it{v_{14}}$ (U$\it{v_{14}}$+TBF) potential. These are to be compared with the empirical values of, respectively, 0.17~fm$^{-3}$, 230$\pm40$~MeV \cite{Khan2012}, and 32$\pm1$~MeV \cite{Baldo2016}. The results indicated that the TBF effect has worked in the direction of increasing the core stiffness of the effective potential. This is also reflected in the binding energy results per nucleon ($E/N$) of neutron as well as beta-stable matter in Fig.~\ref{fig:eqStateBetaStable}. It is clear that the pure neutron matter, with or without TBF, would correspond to a stiffer EOS than the beta-stable matter. Considering the beta-stable two-body results of Bordbar-Riazi and this work together with their slight differences over the range of densities shown in this figure, the inclusion of TNI as compared to TBF would seem to have had a smaller effect on stiffening the potential for densities below about 0.7~fm$^{-3}$. This is reversed for densities above about 1~fm$^{-3}$, where the inclusion of TNI, as compared to TBF, appears to result in significantly higher energies at high density. It could partly be attributed to the exponential construct of the U$\it{v_{14}}$+TNI model, which incorporates higher-than-three-body terms by superposing forces of alternating signs. It could also be attributed to the more complex dependence of $V_{ijk}$ to spin and isospin in U$\it{v_{14}}$+TNI model than the plain central force of $V_{ijk}^R$ in Eq.~\ref{Eq:A2pi}.

It is to be noted that although both of the beta-stable two-body calculations in this figure indicate consistently a smaller stifness as compared with the pure neutron-matter calculations, it would seem not to be the case when the effects of either TNI or TBF are to be added to the corresponding two-body contributions and compared with the neutron-matter results in presence of TBF. The sharp deviation of the beta-stable results plus TNI effect from the neutron-matter results plus TBF effect indicates that, especially at larger densities, the neutron-matter results with TNI inclusion would considerably be stiffer than the ones with TBF inclusion represented in this figure. 

%%%%%%%%%%%%%%%%%%%%%%%%%%%%%%%%%%%%%%%%%%%%%%%%%%%%%%%%%%%%%%%%%%%%%%%%%%%%%%%%%%%

\subsection{Particle densities in beta-stable nuclear matter}
\label{results:2}
Eq.~\ref{asymMatEnergy} allows for calculating the proton fraction under beta-stable equilibrium. Fig.~\ref{fig:particleDensities} compares the electron, muon, and proton densities expected in a beta-stable nuclear matter, assuming that $n\leftrightarrow p+\mu^-$ is energitically allowed above nuclear-matter density at which point the electron chemical potential would surpass the muon mass. Clearly, the muon contribution has ensured significant increase of the proton density, especially at higher nucleonic densities. However, the difference between $E/N$ of the two cases of electrons-only (an equilibrium governed by $n\leftrightarrow p+e^-$ alone) and electrons-plus-muons (an equilibrium governed by both $n\leftrightarrow p+e^-$ and $n\leftrightarrow p+\mu^-$) is not as significant. This difference is estimated in the electrons-only case to increase relative to the electrons-plus-muons case by a maximum of about 7.8\% (U$\it{v_{14}}$ at $\rho=0.67$~fm$^{-3}$) and 2.6\% (U$\it{v_{14}}$+TBF at $\rho=0.59$~fm$^{-3}$). In contrast to the behavior of $E/N$, the proton density is obtained in the electrons-only case to increase with $\rho$ relative to the electrons-plus-muons case, reaching a maximum of about 33\% (U$\it{v_{14}}$) and 34\% (U$\it{v_{14}}$+TBF).

Larger short-range repulsions are expected at high densities, as the short-range repulsion between nucleon pairs that make up isospin singlets dominates the one between isospin triplets~\cite{PandharipandeGarde1972}. Hence, pure neutron matter is to be expected at high enough densities. The reason this is not reflected in Fig.~\ref{fig:particleDensities} data with TBF effect could partly reflect the fact that the central-force repulsion term $V_{ijk}^R$ assumed in the TBF construction does not account for complex spin and isospin dependencies as it should, in order to have a microscopic approach toward the repulsion force. Hence, the particular form of $V_{ijk}^R$ in the TBF construction could be one of the reasons we would not witness the onset of pure neutron matter as we approached toward high densities. As such, further analysis using more realistic nucleon-nucleon models could help pinpoint such problems (e.g. regarding Fig.~\ref{fig:particleDensities}), especially when it concerns the TBF form and the expected effect of $V_{ijk}^R$. 

%%%%%%%%%%%%%%%%%%%%%%%%%%%%%%%%%%%%%%%%%%%%%%%%%%%%%%%%%%%%%%%%%%%%%%%%%%%%%%%%
\begin{figure*}[h]
\includegraphics[width=0.95\textwidth]{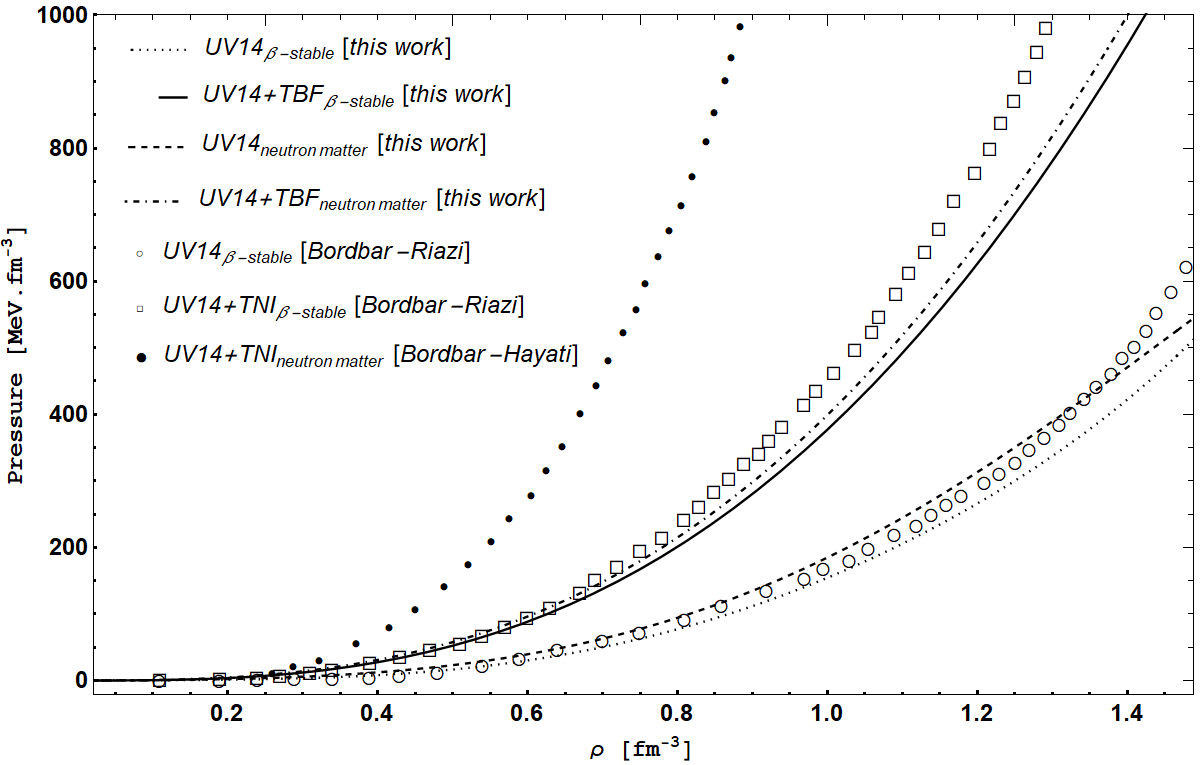}
\caption{\label{fig:pressure}Pressure of beta-stable and neutron matter for different potentials, as a function of nucleon density. The data labeled as Bordbar-Riazi and Bordbar-Hayati were extracted from \cite{BordbarRiazi2002} and \cite{BordbarHayati2006}, respectively.}
\end{figure*}

\subsection{Pressure}
\label{results:3}
Assuming proton and neutron densities of $\rho_{p}$ and $\rho_{n}$ with $\rho=\rho_{p}+\rho_{n}$, the nuclear-matter pressure is obtained as:
\begin{equation}
P={\rho}^2\frac{\partial{E(\rho_{p},\rho_{n})}}{\partial{\rho}}
\end{equation}
Fig.~\ref{fig:pressure} represents our parabolic approximation results for the pressure of the beta-stable and neutron matter with and without TBF. The results indicate generally that accounting for the three-body contribution as TBF or TNI increases the pressure considerably and, in accordance with the results of Fig.~\ref{fig:eqStateBetaStable}, makes the equation of state much stiffer. Considering the effect of three-body interactions on $E/N$ and assuming the overall incompressibility $9{\rho}^2\frac{{\partial}^2{(E/N)}}{\partial{\rho}^2}$, it is to be expected that the three-body effect would add to the incompressibility at a given density -- in agreement with the pressure curves in Fig.~\ref{fig:pressure}. The neutron-matter calculations plus TNI effect predict drastically higher pressures as compared with TBF effect. This is in accordance with the final notes in Sec.~\ref{results:1}, as a result of stiffer potential predicted in the case of TNI inclusion.

\begin{figure*}[h]
\includegraphics[width=0.95\textwidth]{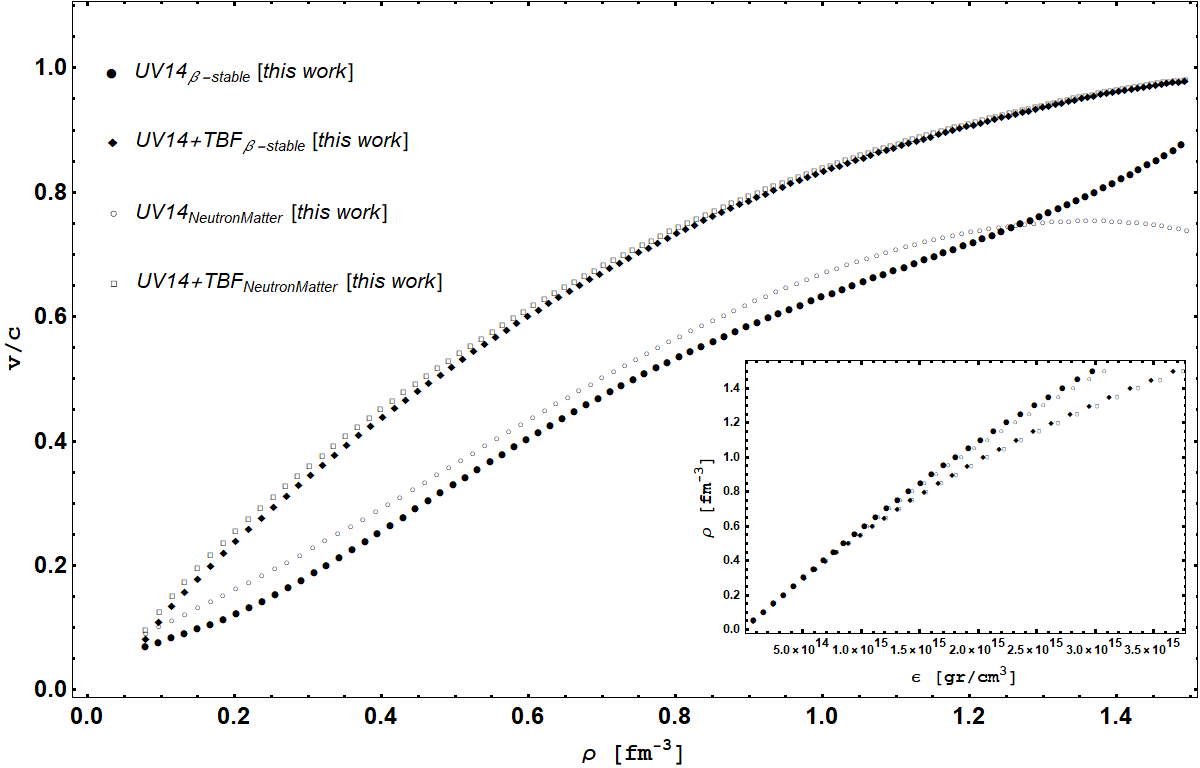}
\caption{\label{fig:soundVelocity}Sound speed in beta-stable and neutron matter with and without TBF. The inset shows how the corresponding nucleon density would change with the mass density.}
\end{figure*}

\begin{figure*}[h]
\includegraphics[width=0.9\textwidth]{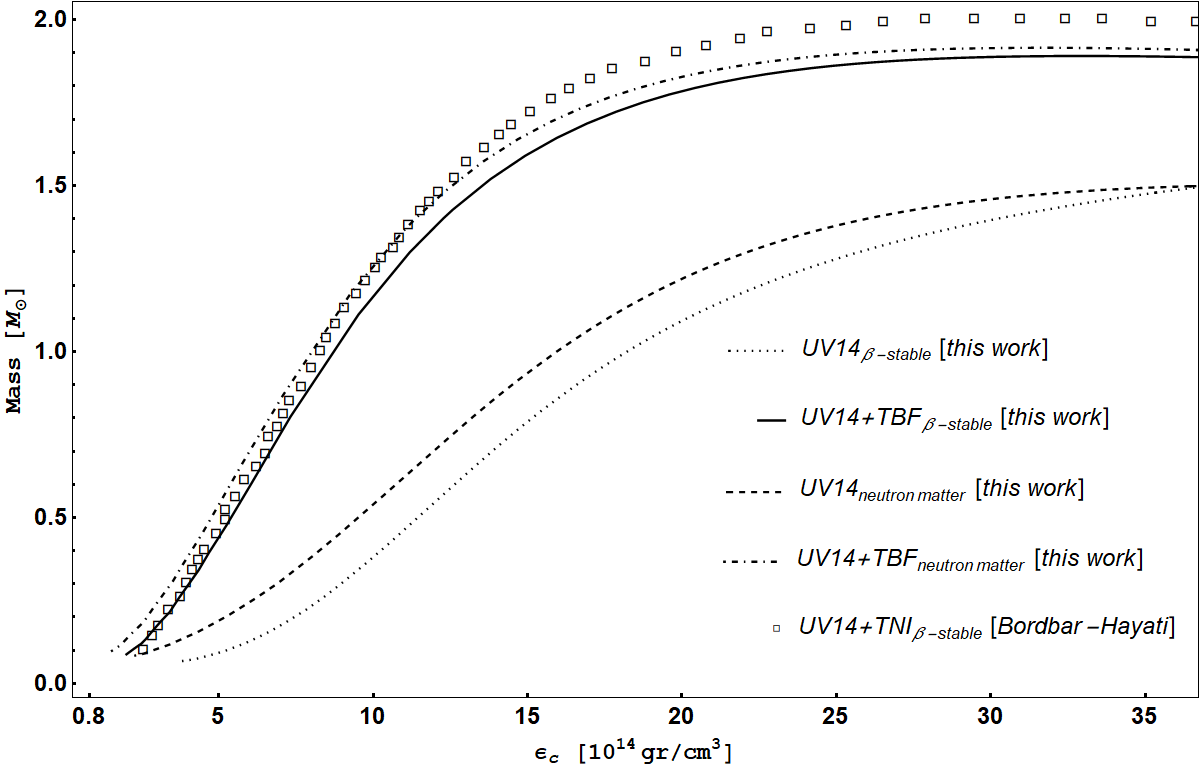}
\caption{\label{fig:Mass_Density}Neutron star's mass in units of the Sun's mass {($M_\odot$)} as a function of its central density {($\epsilon_c$)}.}
\end{figure*}

Given the nuclear-matter pressure, it is interesting to investigate the sound speed in the neutron star's interior as a function of density, $v(\epsilon)=\sqrt{\partial P(\epsilon)/\partial \epsilon}$, which is one of the vital conditions ($v<c$) in addressing the EOS stability~\cite{Abreu2007}. Fig.~\ref{fig:soundVelocity} compares the results for beta-stable and neutron matter, based on U$\it{v_{14}}$ and U$\it{v_{14}}$+TBF potentials. A common feature of the results is that they all respect the causality in that the sound speed does not exceed the speed of light over the investigated densities of up to 1.5~fm$^{-3}$. A clear effect of TBF is the overall increase of the sound speed as compared with the two-nucleon results. This is a reflection of the corresponding pressure results in Fig.~\ref{fig:pressure}, taking into account the small differences of nucleon-density variations against mass density ($\partial \rho/\partial \epsilon$; see the inset of Fig.~\ref{fig:soundVelocity}) as opposed to the sizable differences of pressure variations against nucleon density ($\partial P/\partial \rho$; see Fig.~\ref{fig:pressure}). Indeed, at densities smaller than about 0.5~fm$^{-3}$, it is primarily the rate of pressure change with nucleon density that determines the sound speed in both beta-stable and neutron matter, with and without TBF. Hence, as the pressure variations of various results with density converge at small densities, so does the sound speed values. At ever higher densities, the two factors -- namely, the decrease of $\rho$ variations with $\epsilon$ due to the TBF effect and the increase of pressure variations with $\rho$ -- go against one another to influence the sound speed. Although the two-nucleon results in Fig.~\ref{fig:soundVelocity} appear at high densities to approach the ones with TBF, it is the dominant effect of $\partial P/\partial \rho$ that would guarantee higher sound speeds in presence of TBF as compared with two-nucleon results. In the same line of argument and based on the TBF results in Fig.~\ref{fig:pressure}, higher differences (at ever larger $\rho$ values) of $\partial P/\partial \rho$ between neutron and beta-stable matter seems to have been diminished by the counter-effect of the corresponding $\partial \rho/\partial \epsilon$ results. However, it is not as clear to relate the relative changes of the sound speed results of the two-nucleon cases (beta-stable and neutron) to their corresponding $\partial P/\partial \rho$ behavior in Fig.~\ref{fig:pressure}. This is partly so, since the two pressure slopes do not seem to divert monotonically as a function of $\rho$, which is contrary to what the corresponding TBF results indicate. As such, the two-body neutron matter results above about 1.2~fm$^{-3}$ are suspect -- seen either from the relative change of pressure slope in neutron matter and beta-stable matter or judged certainly from the sound speed in neutron matter which starts to decline unreasonably from about 1.2~fm$^{-3}$ upwards. Though the parabolic approximation has no say in the two-body results of neutron matter -- as opposed to the beta-stable matter -- the sound speed outcomes raise suspicion in neutron matter results at high densities, and this involves the projection of the maximum supportable mass for the neutron star.  

%%%%%%%%%%%%%%%%%%%%%%%%%%%%%%%%%%%%%%%%%%%%%%%%%%%%%%%%%%%%%%%%%%%%%%%%%%%%%
\begin{figure*}[h]
\includegraphics[width=0.9\textwidth]{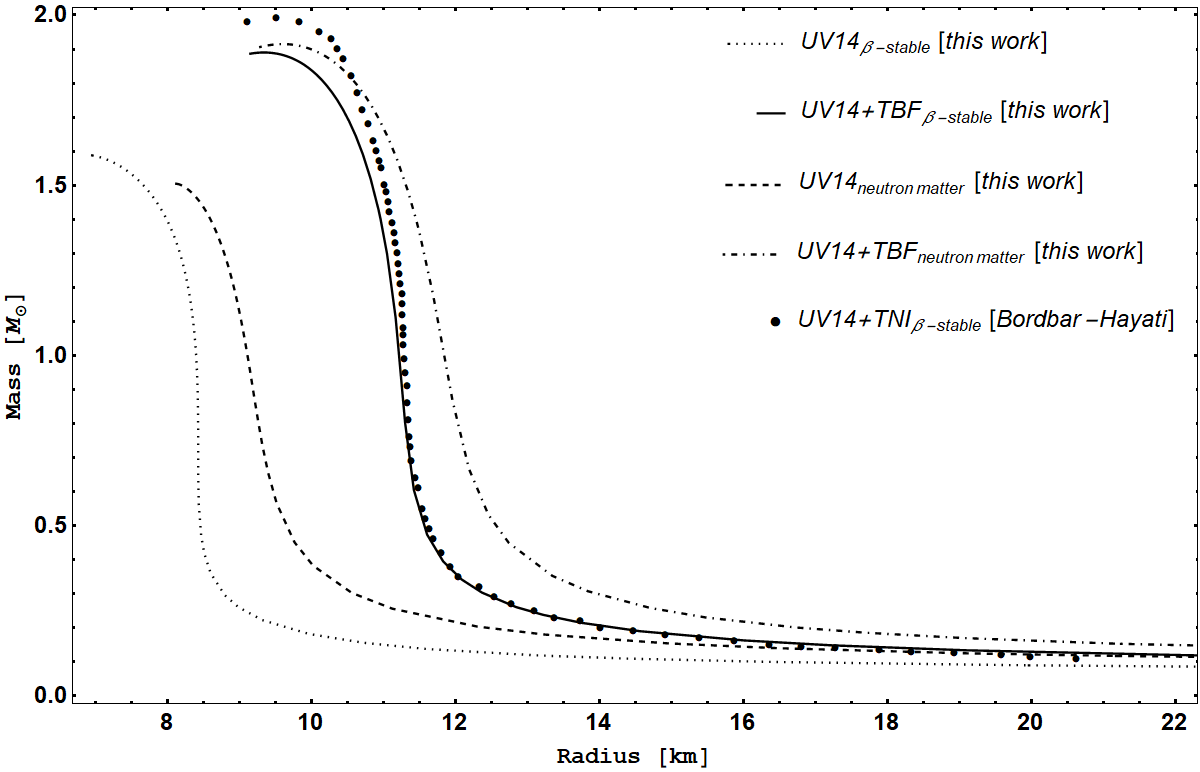}
\caption{\label{fig:Mass_Radius}Neutron star's mass {in units of the Sun's mass ($M_\odot$)}  as a function of its radius.}
\end{figure*}

\subsection{Neutron star's mass, radius, and dynamical stability}
\label{results:4}
Integrating the TOV equation, allows for predicting how the mass or radius of the neutron star would change with its central density and pressure. In our calculations, we have taken into account a crust equation of state before calculating the neutron-star properties. As such, Fig.~\ref{fig:Mass_Density} shows the variation of Neutron star's mass with its central density and Fig.~\ref{fig:Mass_Radius} puts in persective the relation between the mass and radius of a neutron star that is either made purely of neutrons or is in beta-stable equilibrium, assuming a governing U$\it{v_{14}}$ potential in presence and absence of TBF or TNI.

\begin{table*}
\caption{
Different properties of neutron stars calculated in different works, in the absence of magnetic fields. Left column indicates the reference to the work. Next three columns show neutron star's maximum mass ($M$) and its corresponding radius ($R$) and central density. Other columns to the right represent the corresponding Schwarzschild radius $R_{Sch}$, mean density $\overline{\epsilon}$, compactness factor $\sigma$, gravitational redshift $z$, Kretschmann scalar $K$, and the GR compactness limit. Our results constitute the last four rows. Here, G, c, and $M_{\odot}$ refer to the gravitational constant, light speed, and the Sun's mass, respectively.}
%\begin{center}
{
\begin{tabular}{|c|ccc|cccccc|}
\hline
{\footnotesize Ref.}& $M$ & $R$ & $\epsilon_c/10^{15}$ &$R_{Sch}$ & $\overline{\epsilon}/10^{15}$ & $\sigma$ & $z$ & $K/10^{-7}$ &$\frac{4c^2R}{9G}$\\
& $[M_{\odot}]$ & $[km]$ & $[g/cm^3]$ & $[km]$ & $[g/cm^3]$ & & & $[1/m^{2}]$ & $[M_{\odot}]$\\
\hline
{\footnotesize \cite{EslamPanahBordbar2017}}& 1.68 & 8.42 & - & 4.96 & 1.34 & 0.59 & 0.56 & 0.29 & 2.53 \\
{\footnotesize \cite{BordbarRezaei2013_2}}& 1.69 & 8.59 & - & 4.99 & 1.27 & 0.58 & 0.54 & 0.27 & 2.58 \\
{\footnotesize \cite{BordbarKarami2022}}& 1.68 & 9.00 & - & 4.96 & 1.09 & 0.55 & 0.49 & 0.23 & 2.71 \\
{\tiny $\beta$-stable matter:}& & & & & & & & & \\
{\footnotesize U$\it{v_{14}}$}& 1.59 & 6.96 & 5.37 & 4.70 & 2.24 & 0.67 & 0.75 & 0.48 & 2.09 \\
{\footnotesize U$\it{v_{14}}$+TBF}& 1.89 & 9.36 & 3.26 & 5.58 & 1.09 & 0.60 & 0.57 & 0.23 & 2.82 \\
{\tiny neutron matter:}& & & & & & & & & \\
{\footnotesize U$\it{v_{14}}$}& 1.50 & 8.13 & 4.38 & 4.43 & 1.32 & 0.54 & 0.48 & 0.28 & 2.45 \\
{\footnotesize U$\it{v_{14}}$+TBF}& 1.91 & 9.59 & 3.19 & 5.64 & 1.03 & 0.59 & 0.56 & 0.22 & 2.88 \\
\hline
\end{tabular} \label{tab:table1}
}
%\end{center}
\end{table*}

\begin{figure*}[h]
\includegraphics[width=0.9\textwidth]{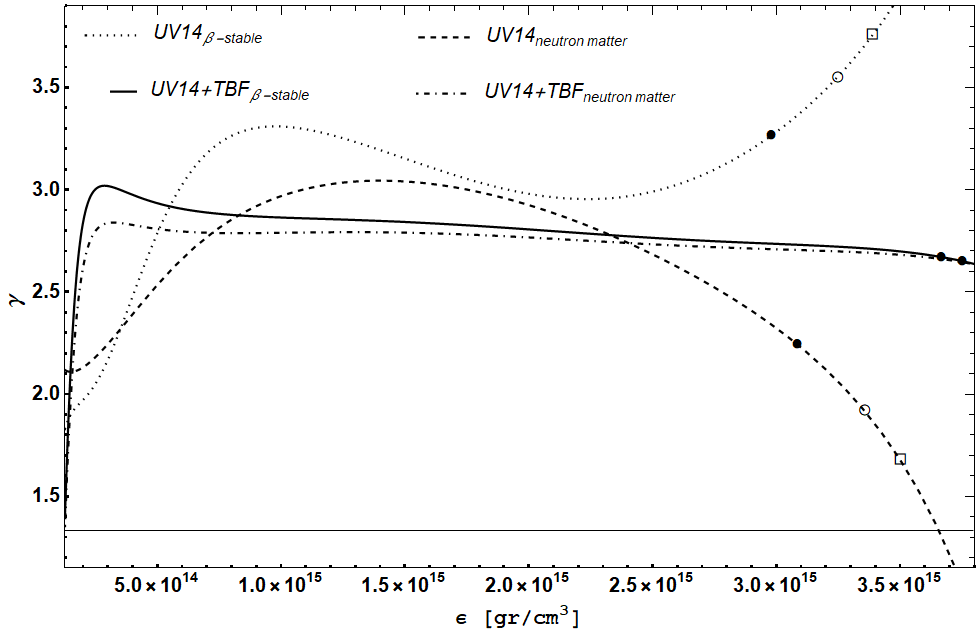}
\caption{\label{fig:AdiabaticIndex}Adiabatic index versus density, for $\rho>0.07$~fm$^{-3}$. The full circles, empty circles, and empty squares on each curve correspond to $\rho$=1.5, 1.6 and 1.65~fm$^{-3}$, respectively.}
\end{figure*}

Along with other calculations, Table~\ref{tab:table1} shows our calculations for the maximum mass and the corresponding radius of neutron stars -- under beta-stability equilibrium as well as made of pure neutron matter -- based on which the values of few characteristic parameters were obtained. These include the Schwarzschild radius $R_{Sch}=2GM/c^2$, mean density $\overline{\epsilon}=3M/4\pi R^3$, compactness factor $\sigma=R_{Sch}/R$, gravitational redshift $z=\frac{1}{\sqrt{1-2GM/c^2R}}-1$, Kretschmann scalar $K=4\sqrt{3}GM/c^2R^3$~\cite{Psaltis2008,Eksi2014}, and Buchdahl-Bondi upper mass limit $M_{max}\leq4c^2R/9G$~\cite{Buchdahl1959,Bondi1964,Buchdahl1966}. Since our results for the radius of the neutron star are more than the maximum Schwarzschild radii, associated with their respective maximum mass, none of our hypothesized neutron stars made of either pure neutron or beta-stable matter (with and without TBF) are expected to end up with a black hole. In general, the TBF effect has translated into an increased $R_{Sch}$, which is clearly what we expect also from the neutron star's maximum mass. Unlike the expected increase in both of the maximum mass and the corresponding neutron star's volume due to the TBF effect, the resulting average density appears to shrink relatively ($\Delta\overline{\epsilon}/\overline{\epsilon}$) by about 54\% and 22\% in the case of beta-stable and neutron matter, respectively. Hence, a lower average density due to TBF together with the fact that the overall pressure increases due to TBF (see Fig.~\ref{fig:pressure}) means that as the neutron star's overall pressure increases due to the TBF effect, so does the average inter-nucleon distance. Thus, it is not surprising that given a neutron star's mass, the TBF effect as compared to lack thereof have resulted in a larger radius (see Fig.~\ref{fig:Mass_Radius}). Similar situation arises either in the presence or absence of TBF, by considering the overall pressure of the pure neutron matter which is higher than the beta-stable matter, contrary to the resulting average density of a neutron star purely made of neutrons which is smaller than its average density in beta-stability equilibrium. The compactness factor which is a measure of the gravity strength is proportional to $M/R$, approximately resembling the behavior of the gravitational redshift as a function of radius. The Kretschmann scalar $K$ is a measure of the neutron star's curvature at its surface and, due to an extra dependence on $R^{-2}$, resembles $\sigma$ or $z$ to a lesser degree so that its values for the neutron matter and corresponding to $M_{max}$ have appeared in different order than $\sigma$ or $z$ values. The numbers in the right column show the general-relativity compactness limit which is the upper mass limit for a static spherical neutron star of constant density. The fact that the maximum-mass values are obtained to be smaller than Buchdahl-Bondi limit is another indication that the hypothesized neutron stars in this work (made of pure neutron or beta-stable matter bound by U$v_{14}$ potential, in presence or absence of TBF, and governed by the TOV equation) would not turn into black hole. The dynamical stability, which was defined by Chandrasekhar~\cite{Chandrasekhar1964}, is a concept introduced to check the neutron star's stability against infinitesimal radial adiabatic perturbations and is fulfilled so long as the adiabatic index $\gamma=\frac{\epsilon c^2+P}{c^2 P}\frac{dP}{d\epsilon}>4/3$, which has been checked for many astrophysical cases including~\cite{Kuntsem1988,Mak2013,Kalam2015}. Fig.~\ref{fig:AdiabaticIndex} represents our results for the adiabatic index as a function of density, showing that the dynamical-stability condition is satisfied for the hypothesized neutron stars studied over $\rho\leq 1.5$~fm$^{-3}$.

Table~\ref{tab:table2} puts the measured mass and in some cases -- where the measurement of radius succeeded through complex procedures involved in observation -- the radius of a few neutron stars into persective. The masses span over about one to two times the mass of the Sun. Given a measured mass, the data on the right side demonstrate our calculations for radius. The calculations correspond to pure neutron as well as beta-stable matter, in which electrons and muons both have contributed to hold up the equilibrium. There are fields that were left empty, since our results would not predict masses as large as the measured ones. Incidentally, all our results are compatible with the observed masses smaller than about 1.50~$M_{\odot}$ (see Table~\ref{tab:table1}, left column) in that they could work out a radius corresponding to the observed mass. But, for masses above 1.59~$M_{\odot}$, they would deliver a radius only when TBF is accounted for; hence, they could only amount to a radius for two of the observed masses (above 1.59~$M_{\odot}$) in Table~\ref{tab:table2}. Parenthetically, our pure-neutron and beta-stable results both agree -- in present of TBF -- with the measured radius of VelaX-1~\cite{Rawls2011} within the uncertainties. The reason our calculations could not work out a radius for masses as high as about 2~$M_{\odot}$ could partly be due to the possibility of quark-hadron-phase existence within the neutron star, in which case our model of a neutron star -- purely made of nucleonic matter -- would break down. Indeed, there are studies on PSRJ0348+0432 and PSRJ1614-2230 (see Table~\ref{tab:table2}) arguing that there may exist a region of quark-hybrid matter within their core~\cite{Orsaria2013,Orsaria2014}, or that compact stars with masses close to 2~$M_{\odot}$ (like the three cases in Table~\ref{tab:table2}), are compatible with deconfined quark matter presence at their core~\cite{Lastowiecki2015}.

\begin{table*}
\caption{
Measured mass and radius of few neutron stars through observation. Right columns: our estimates for the radius corresponding to the measured mass.}
%\begin{center}
{
\begin{tabular}{ccc|cccc}
\toprule
\multicolumn{3}{c}{Observation} &
\multicolumn{4}{|c}{Calculated $R~[km]$} \\
& & &
\multicolumn{2}{c}{beta-stable} & \multicolumn{2}{c}{neutron-matter} \\
\hline
Name~[Ref.] & $M~[M_{\odot}]$ & $R~[km]$ & U$\it{v_{14}}$ & U$\it{v_{14}}$+TBF & U$\it{v_{14}}$ & U$\it{v_{14}}$+TBF \\
\midrule
{\footnotesize SMC X-1~\cite{vanDerMeer2005}} & $1.05\pm 0.09$ & - & 8.39 & 11.20 & 9.08 & 11.80 \\
{\footnotesize Cen X-3~\cite{vanDerMeer2005}} & $1.24\pm 0.24$ & - & 8.25 & 11.09 & 8.88 & 11.63 \\
{\footnotesize LMC X-4~\cite{vanDerMeer2005}} & $1.31\pm 0.14$ & - & 8.16 & 11.04 & 8.77 & 11.55 \\
{\footnotesize V395 CAR/2S 0921C630~\cite{Steeghs2007}} & $1.44\pm 0.10$ & - & 7.89 & 10.92 & 8.49 & 11.40 \\
{\footnotesize PSRJ0740+6620~\cite{Cromartie2020}} & $2.10$ & $12\pm 2$ & - & - & - & - \\
{\footnotesize PSRJ0348+0432~\cite{Antoniadis2013}} & $2.01$ & $13\pm 2$ & - & - & - & - \\
{\footnotesize PSRJ1614-2230~\cite{Demorest2010}} & $1.97$ & $12\pm 2$ & - & - & - & - \\
{\footnotesize VelaX-1~\cite{Rawls2011}} & $1.80$ & $11\pm 2$ & - & 10.19 & - & 10.62 \\
{\footnotesize 4U1608-52~\cite{Guver2010}} & $1.74$ & $9\pm 1$ & - & 10.39 & - & 10.82 \\
\bottomrule
\end{tabular} \label{tab:table2}
}
%\end{center}
\end{table*}
%%%%%%%%%%%%%%%%%%%%%%%%%%%%%%%%%%%%%%%%%%%%%%%%%%%%%%%%%%%%%%%%%%%%%%%%%%%%%%

\section{Summary and conclusions}
\label{sec:conclusions}

Performing calculations for the asymmetric nuclear matter with the help of parabolic approximation and U$v_{14}$ potential, we have investigated the effect of a newly constructed phenomenological three-nucleon force which was constructed exploiting two-body correlations -- derived using the LOCV method and the concept of three-body radial distribution function -- the details of which were discussed in~\cite{MoeiniBordbar2022}. Applying the method to the specific cases of pure neutron and beta-stable matter allowed us to assess the TBF effect on various particle densities as well as the bulk properties of neutron stars. These included the influence of TBF on the sound speed and adiabatic index as well as how the mass and radius of the neutron star would change with its central density and pressure, and as a result what would be its maximum mass and corresponding radius. Obtaining the neutron star's maximum mass has a special importance in that it indicates that the degeneracy pressure of nucleons would be enough not to allow the neutron stars with $M\leq M_{max}$ to turn into black holes~\cite{Shapiro1983}.

The TBF effect seemed to have been in the direction of increasing the neutron star's maximum mass and decreasing the central density associated with maximum mass. Investigating the dependence of the radius on the central density showed, generally, that the radius would decrease as the central density increases. More specifically, at small values of central density or pressure, the radius would experience a relatively sharp drop as the central density or pressure grows. Beyond a certain central pressure or dencity (around $5\times10^{14}$~g/cm$^3$ with TBF and $9\times10^{14}$~g/cm$^3$ without TBF), there appears a drastic change where the radius would not shrink as sharp. Our hypothesized neutron star, constructed using U$v_{14}$ potential+TBF+parabolic approximation+TOV equation, could predict a radius for all the observed masses below 1.89~$M_{\odot}$ (beta-stability results) or 1.91~$M_{\odot}$ (neutron-matter results). In the observation case of VelaX-1~\cite{Rawls2011}, both of the radius results (neutron and beta-stable matter) agreed with the observed one within the reported uncertainties.

Knowing that there are inherent problems regarding the U$v_{14}$ potential and in order to study the significance of the proposed TBF and its implications for the beta-stable matter and neutron star's stability, we are encouraged to further investigate the prospects of the TBF effect in conjunction with more realistic nucleon-nucleon models constructed on the basis of large-scale scattering databases at intermediate energies.

\section*{Acknowledgments}
We wish to thank the Shiraz University Research Council.
\\
\\
\noindent{Data Availability Statement:} No Data associated in the manuscript

\end{document}